# ATOMIC NUCLEUS AS A LABORATORY FOR FUNDAMENTAL PROCESSES*

J. Dobaczewski

Institute of Theoretical Physics, Warsaw University
ul. Hoża 69, PL-00681 Warsaw, Poland

Ladies and Gentlemen,

The 28th Mazurian Lakes Conference is about to end. We have spent here seven days, listened to many lectures, and had numerous discussions. Only one lecture, this one, separates you from lunch, concert, and conference dinner. So the intermediate future looks bright. However, as for the immediate future, it is somehow grimmer. You have in front of you a completely inexperienced speaker who has never given any concluding remarks talk at any conference. This makes me really nervous, and I can tell you, this should make you nervous too. I still keep asking myself what on earth I am going to tell you?

When preparing this talk I went to library to look for useful patterns and I realized with dire fear that such talks are usually funny, and I think they are supposed to be funny – everybody thinks they should be funny. The problem is that I am not too funny a kind of person. I am very bad in telling jokes, especially in public. I therefore specifically asked our Conference Chairmen Ziemek Sujkowski for a license not to by funny, and he granted it to me. This took out some of the stress, but the share that was left is still paralyzing me, indeed. I also asked Ziemek for a confirmation in writing that I would be delivering concluding remarks and not a summary talk. So I am under no obligation to mention every talk we heard, sorry folks.

From the literature and from my own experience I know, of course, that there do exist easy methods to give concluding remarks. The first one is to say: *You guys have had here a nice conference, but now, let me tell you about really interesting stuff*, and then go on with the talk on my own research. The second one is applicable in case my collaborator would have already

---

* Concluding remarks presented at the XXVIII Mazurian Lakes Conference on Physics, Krzyże, Poland, August 31 – September 7, 2003

given a talk on our common research. Then I could say: *I do not think everyone really grasped the essence of our research*, and then go on with delivering my collaborator's talk once again. I have witnessed these two methods in action with my own wide opened eyes. I can tell you – the effect they exert on the audience is truly staggering. Both of these easy methods also provide a non-negligible perk: I would have never again been asked to give the concluding remarks.

This talk is the first one for me also from another point of view: I have never given a talk projected from a computer. This fashion is now spreading like a forest fire, and I do not think there is any real possibility to stop it. During this conference more than half of the talks were delivered in this way, and I did not have any choice; I had to collect files from some speakers (thank you!), and foils from some others (thank you!). Thus I will show you some things from the computer and some from the projector. Two screens shown at the same time will annoy you, and I am certainly up for an even bigger disaster.

I do not like computer presentations, because the present-day technology is not up to my expectations yet. First of all, there are always small technical problems (cable connections fail or resolutions do not match), which require shorter or longer breaks in the session, and make the audience laugh. Second, existing software delivers presentations that are essentially linear; one screen after another. So if I am going to skip a part of my presentation, or go back a few screens, I have to go click, click, click, and everyone sees things that I want to skip. This looks odd. I think that such a presentation could run smoothly only if I had a possibility to see one screen on my laptop, and to project a different screen for the public. Then, I could steer my presentation the way I want without letting the audience know.[1]

Many speakers using computers fall in the trap of showing cheep animations of objects, which arrive erratically from right, left, top, or bottom, without any visible purpose or sense. This always distracts my attention from the subject matter of the talk, and I have a feeling that I am not alone. On the other hand, the real power of computer presentations lies in showing animations that do make sense. During this conference Matthias Liebendör­fer and Terry Awes showed us useful animations. They were really good. Matthias not only performed prime quality general-relativity calculations of exploding supernovae, but also showed us movies of how many, and from which layer the neutrinos are emitted. In his case the evolution in time was illustrated by animation. Terry used animated graphics to show us dependence of results on a "third variable". Functions of transverse momentum were shown for several centralities in an animation that perfectly illustrated his physics point. I think that a good and useful computer animation is invaluable. A couple of decades ago we moved from communicating in words (blackboard) to communicating in pictures (transparencies), and obviously one picture speaks a thousand words. Now it is time to start communicating in movies; one movie certainly shines a thousand pictures.

---

[1] After the talk I was told that there are ways to set up the computer for two screens. Somehow, nobody uses this option yet.

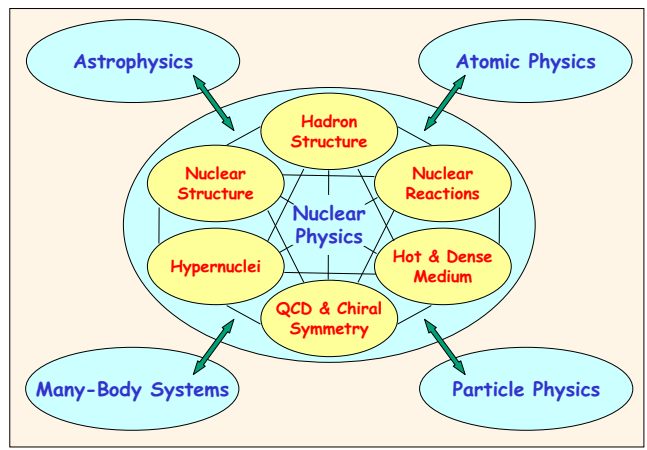

Fig. 1. Diagram shows subdomains constituting nuclear physics of the present day, and main links between nuclear physics and other domains of physics.

Concluding remarks after a scientific conference are a perfect opportunity to reflect about a general structure of the domain that has been discussed. In our case it is especially appropriate, because our domain, the nuclear physics, underwent significant changes during the past ten or fifteen years. It is not any more associated only with low-energy properties and reactions of nuclei, but includes several other aspects, previously associated with particle physics. One can now say that every physical phenomenon that is rooted in the Quantum Chromodynamics (QCD) of *uds* quarks belongs to nuclear physics. This change has not been mandated by a decree or a decision of a commission; it happened naturally as a result of how physicist doing physics call what they do. It is enough to check out sections and papers in a major nuclear physics journal, like those of *Physical Review C* or *Nuclear Physics A*, or peruse proceedings of a major nuclear physics conference, like those of the 2001 *International Nuclear Physics Conference* in Berkeley, or browse the Nuclear Theory archive at *http://arxiv.org/archive/nucl-th*, to see how nuclear physicists presently define the nuclear physics.

Figure 1 shows a schematic diagram of the present-day subdomains of nuclear physics along with the links that it has with other domains. Apart from the traditional physics of nuclei, i.e., the nuclear structure and nuclear reactions, and of hypernuclei, it also contains (i) physics and structure of hadrons, including the structure of the vacuum, mesons, and nucleons, as well as their excited states, (ii) physics of nucleonic matter at high densities and/or temperatures, and (iii) basic properties and symmetries of QCD in its low-energy, non-perturbative regime, including the chiral symmetry breaking. Interconnections between the subdomains of nuclear physics are very strong.

Nuclear physics has also particularly strong links with astrophysics. This is obvious, because stellar objects are powered by nuclear energy, and live their life and demise according to nuclear rules. Particle physics now deals with QCD for heavy quarks, electroweak interactions, unification schemes and cosmology, and often represents the high-energy end of nuclear physics. On the other end, nuclear physics is strongly connected with the atomic physics, where it constitutes an important laboratory for fundamental interactions and precise measurements. Furthermore, many-body aspects, which are always present in nuclear systems, find their mirror images in many other physical objects like atom drops and clouds, metal clusters and grains, or quantum dots and wires.

This conference was focused on hypernuclei, nuclear structure, and hot and dense nucleonic medium. The nuclear reactions and QCD aspects were also nicely covered. Only the hadron structure was not very much represented, so the recent advances in studies of internal structure of nucleons are left for a future meeting in Krzyże. The astrophysics and nuclear astrophysics were both very well represented, while from the particle physics, the neutrino physics was at the focus. This year, the atomic physics and many-body aspects of other system were not discussed; again these are perfect candidates to become the main topics after two years.

Visibly, the Mazurian Lakes meeting has evolved from a topical school on specific aspects of nuclear structure to a general-interest nuclear physics conference. I think this is good. There is certainly a great need for such a kind of conference in Poland, because the changes that have occurred in our domain, did not yet fully occurred in our colleagues' minds. Moreover, there is a great need for an improved communication among nuclear physicist. It is out of question that any single individual might actively work in all, or many subdomains of nuclear physics – the field is simply too large, and the active work means (should mean) a certain level of expertise. However, several common aspects of our understanding of physical reality beautifully unite the domain, and we have to mutually learn, admire, and respect the way we do the physics we like. The conference fulfilled this requirement very well.

A general-interest conference requires a larger effort from speakers, who have to refrain from focussing on details, and convey a wider picture to potentially unprepared audience. Most speakers did very well in this respect, but a few did not. I was especially saddened by several "too-many-details" presentations of young speakers, who probably were not properly advised by their supervisors. This reminds me about a commercial for a public-transit company, which I saw on a bus in Philadelphia. It said "YRU DRIVING?". As a non-American I had at first a hard time to figure out what does this suppose to mean? Only after a little thinking, it occurred to me that the real question was "Why are you driving?". And it made perfect sense: if a driver does not know why he/she is driving, he/she should have probably taken a bus. So before preparing your transparencies, or a computer file, it is always healthy to ask yourself questions like "YRU doing this research?" and "What is your bottom line?". If you spend 60% of your talk on answering these questions, and only the remaining 40% on describing your apparatus, calibration, diagonalization. and/or convergence, your talk will be fine.

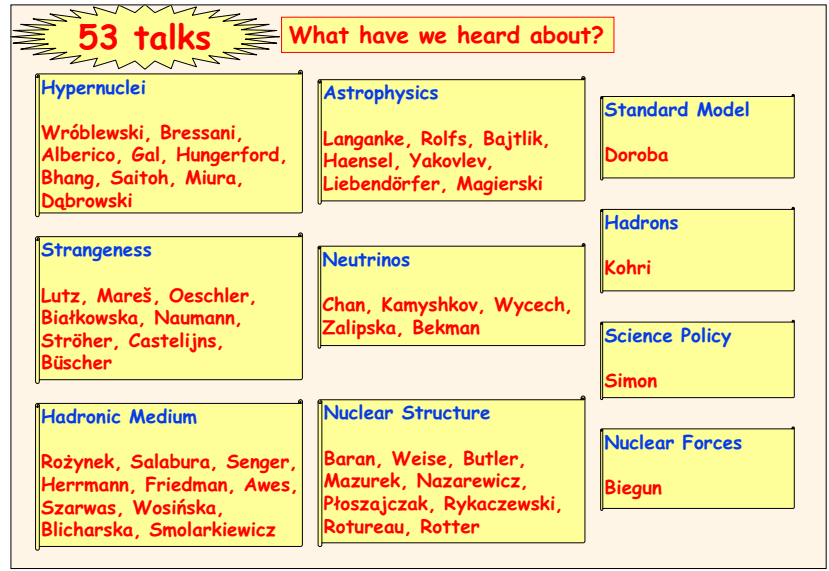

Fig. 2. Diagram lists names of speakers according to main subjects that have been discussed during the conference.

I think it is already great time to start speaking about the contents of talks presented during the conference. In the diagram shown in Fig. 2, I tried to group names of speakers into thematic chapters. Within each chapter, the names are listed according to the order in which the talks were delivered. I was particularly pleased to be able to put Wolfram Weise within the chapter of "Nuclear Structure"; Wolfram – welcome in good company! Of course, many of the talks could be assigned to several chapters simultaneously, e.g., Wolfram's talk also pertains to "Hadronic Medium" and "QCD". A network of such connections would certainly cover the whole diagram with arrows pointing in all possible directions.

I am certainly not enough qualified to be able to expertly talk about all the chapters. However, I have certainly learned some interesting new things (at least, new for me) during this conference. In what follows, I try to give you some glimpses of what caught my eye and ear, and what you have managed to teach me.

One of the main themes discussed during the conference was motivated by the 50th anniversary of discovery of hypernuclei by Danysz and Pniewski. The history of how it all begun and developed was presented in a fascinating introductory lecture by Andrzej Wróblewski. Numerous single-$\Lambda$ and a few double-$\Lambda$ hypernuclei have been up to now discovered and studied experimentally. However, as compared to the enormous amount of data and knowledge accumulated over the years for normal nuclei, relatively little is still known about hypernuclei. The reasons are, of course, in difficulties related to produce them and study within the narrow window of their lifetimes. For example, Yusuke Miura told us about the recent important progress in the $\gamma$-spectroscopic studies of hypernuclei, but we are all aware of how much more extensive are similar studies of usual nuclei. Similarly, basic strengths of the $\Lambda\Lambda$ versus $N\Lambda$ interactions, discussed by Abraham Gal, are still debated, and conclusions are drawn from the simplest binding energy differences between single-$\Lambda$ and double-$\Lambda$ hypernuclei. This stage has already been long time ago passed for nuclei.

Very little is known about the three-body forces ($NN\Lambda$ and $N\Lambda\Lambda$) in hypernuclei. Paweł Haensel gave us good arguments that such forces could be even more important for hyperons than they are for nucleons, because the $\Sigma$-$\Lambda$ excitation energy is much smaller than that for $\Omega$-$N$. Paweł also speculated on how such forces may influence a hypothetical hyperon core of a neutron star.

Wanda Alberico and Hyoung Chan Bhang discussed weak non-mesonic decay rates of hypernuclei. The long-standing and still debated issue here is the so-called $\Gamma_n/\Gamma_p$ puzzle, i.e., the fact that the observed widths for $\Lambda + n \to n + n$ decays seem to be much larger, relatively to $\Lambda + p \to n + p$, than those obtained theoretically. A recent progress, both in theory and in experiment, seems to yield more convergent results. The Physical Review Letters article published on the first Tuesday after the conference presents similar conclusions. The whole issue is very much complicated by the in-medium and final-state effects, while the extraction of pure non-mesonic

7decays rates of hyperons is crucial for our knowledge of weak flavor-changing baryon-baryon interactions.

Experimental attempts to study physics of strange quarks were discussed this morning, and will also be addressed during the coming *VI ANKE Workshop*, which follows our conference. Earlier, Helmut Oeschler told us about experimental studies that aim at seeing the quark-gluon plasma (QGP) through the lens of strangeness production. Indeed, if a flavor-equilibrated fireball of quark matter was produced in energetic nucleus-nucleus collisions, the production of strange and non-strange particles should have been comparable. This can be quantified in the form of the Wróblewski factor, $\lambda_S{=}2s\bar{s}/(u\bar{u}+d\bar{d})$, and studied through strange mesons, strange baryons, or hidden strangeness production. However, within the statistical model, experimental maxima of $\lambda_S$, that reach values around 0.6, can be explained by a kinematical cut through the $T$–$\mu_B$ plane, and thus need not signal the presence of the QGP. This conclusion can be, of course, subjected to usual questions about the validity of statistical model, which can be rigorously applied only to infinite (or at least very large) and well-equilibrated systems.

Peter Senger gave us a very nice account of the future international accelerator facility at GSI. He described the main scientific directions the facility will aim at, namely, the hadron spectroscopy, structure of nuclei far from stability, high energy density in matter, and compressed baryonic matter – a truly nuclear physics facility! This is really a fantastic project that will give us tremendous amount of data and boost our knowledge of nuclear systems. We all hope that the famous missing 25% of European funding will be found, and that the project will go ahead full steam as rapidly as possible. The missing funds will not come from Poland, though. This statement by no means reflects our evaluation of the scientific merit of the project, not at all. Unfortunately, it reflects the criminally low level of the science funding in our country, which makes that on May 1, 2004 we may nominatively join the European Union, and factually (in science) we may join Africa.

How to measure the in-medium hadron mass was discussed by Piotr Salabura. He convinced us that the dielectron two-body decays into $e^+e^-$ pairs can be used to directly measure the invariant mass of a decaying hadron, because leptons traveling through nuclear medium are not perturbed in their final states. Also the Dalitz decays, in which the $e^+e^-$ pair is accompanied by a hadron, can also be well used. The problem here lies in a necessity to disentangle contributions coming from various decaying hadrons. Piotr described the HADES experiment at GSI that aims at this kind of studies.

A very interesting idea to study the equilibration process in nucleus-nucleus collisions was discussed by Norbert Herrmann. Namely, by using projectiles and targets that have different isospin compositions one can, in a sense, tag the nucleons that originate from the projectile or target, and see if they bounce of each other, equilibrate, or pass each other. Experimental data obtained at SIS energies of 100–200 $A$ MeV clearly indicate that the colliding systems are never fully stopped and that the transparency increases with incident energy. Thus, at least in this case, we have a clear experimental indication that the systems are not really equilibrated.

At a completely different scale of energies, $\sqrt{s_{\text{NN}}}$=130 and 200 GeV, studied at RHIC, Therry Awes compared the nuclear modification factors $R_{AA}$ (for Au+Au collisions) and $R_{dA}$ (for d+Au collisions) in order to pin down the so-called jet quenching effect. At these energies, jets of particles are produced when the projectile and target quarks collide and create flux tubes that then break into colorless hadrons. The nuclear modification factor gives a rate of production of a given hadron in a nucleus-nucleus collision, relative to that for the proton-proton collision. It is meant to tell us how much the medium, in which the particle is produced, influences the observed outflow of particles after the collision. For a fairly wide region of transverse momenta, values of $R_{dA}$ are close to one, while those of $R_{AA}$ are suppressed to about 0.3. This is a strong indication that a new kind of medium (possibly the QGP) is created in the Au+Au collisions.

Within the chapter of astrophysics, Karlheinz Langanke told us how strongly properties of stellar objects may depend on detailed nuclear structure properties. First of all, he showed us very impressive results of large-scale shell-model calculations for the Gamow-Teller strength distributions. These calculations agree with the newly measured (with a 100 keV resolution) data incredibly well, which shows that the low-energy nuclear properties are well under control by using two-body interactions in a restricted valence space. Second, he showed us how similar calculations, performed in heavier nuclei within the Monte-Carlo shell model, modify simplistic electron capture rates, which were up to now used to model supernovae explosions. The effect is truly dramatic, because the capture on nuclei now turns out to be more important than the capture on protons, assumed previously. Karlheinz was visibly disappointed that after he performed a decent nuclear structure calculation his supernovae still did not wish to explode. However, to perform a decent job never hurts, and the frontier between nuclear physics and astrophysics is especially prone to simplistic treatments that have to be systematically eradicated one by one.

On the frontier between astrophysics and neutrino physics, we had two nice talks about stellar objects viewed by looking at their neutrino emission. Matthias Liebendörfer investigated time and energy characteristic of neutrinos emitted during supernova explosion. If one of these happens again nearby, we may learn from the observed neutrino flux how such an event proceeds, provided we have a good model at hand.

Dima Yakovlev told us that a hot neutron star cools down mainly due to neutrino emission. This is so, because the neutron decay followed by the inverse process of electron–proton recombination, results in an emission of one neutrino and one antineutrino. Since these processes crucially depend on details of occupations near the Fermi surfaces, and thus depend on correlations, the proton and neutron pairing may strongly influence the rate of cooling. Experimental data seem to suggest that the proton pairing may be preferred over the neutron pairing, which is a surprising conclusion. Since the pairing in neutron stars was up to now freely studied (in the absence of data), a definite experimental constraint of some kind would be very much welcome.

Among several talks on the neutrino physics, Yuri Kamyshkov and Joanna Zalipska presented experimental studies performed at KamLAND and Super-K facilities, respectively. The neutrino oscillation phenomenon is now nicely confirmed, both by the reactor electron antineutrino and muon neutrino observations, and the results converge towards the so-called large mixing angle (LMA) solution for the neutrino mass difference and mixing angle.

Now we arrive at the chapter of nuclear structure, and only here my talk really begins, because on this subject I can talk for hours and hours. But first, let me describe the results presented by Wolfram Weise, which are situated on the triple frontier between QCD, hadronic medium, and nuclear structure. Recent developments in this field are really fascinating, because we may witness the birth of first-principle derivations of nuclear forces. This is being done in kind of a perfidious way, by telling us that we should simply not use nuclear forces at all! (I think this is probably an extreme point of view.) However, it turns out that by applying ideas based on the chiral symmetry breaking, chiral condensate, and effective field theory (EFT) we may describe nucleon-nucleon (NN) scattering and finite nuclei almost directly from low-energy QCD considerations.

This is done by postulating the chiral Lagrangian of nucleons and pions, which is next supplemented by symmetry-dictated contact terms that are supposed to describe all unresolved high-energy effects. After performing a systematic classification of these contact terms, according to the counting rules of the EFT, and by going to the so-called next-to-next-to-next-to-leading order expansion (N$^3$LO), one is able to properly describe the NN phase shifts up to about 300 MeV (D.R. Entem & R. Machleidt, nucl-th/0304018), with quality similar to that achieved for the best NN potentials. Such a result shows that the short-distance NN repulsion need not be modelled by any kind of hard-core potential, or heavy-meson exchange potential, but is a generic feature of these unresolved high-energy effects. Similarly, as Wolfram described in his talk, one may perform in-medium chiral calculations and derive the energy density functional, which within the relativistic-mean-field approximation is directly applicable to finite nuclei. At the expense of fitting one parameter, the EFT cut-off energy, one can obtain correct saturation energy, saturation density, and symmetry energy. From there, standard nuclear structure calculations lead to describing nuclear masses (only of $N=Z$ nuclei, at present) with a precision of about 1 MeV.

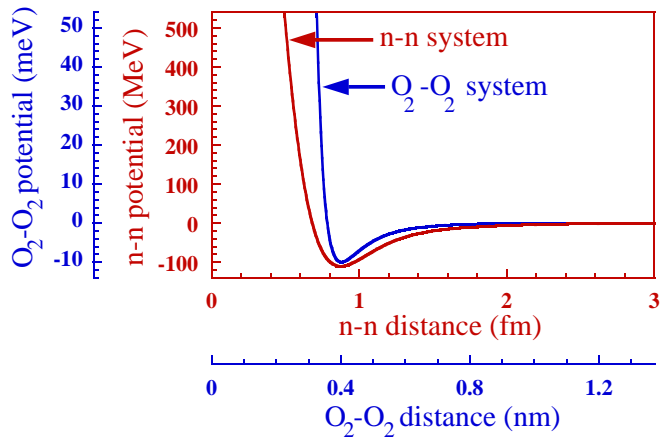

Fig. 3. The $^1S_0$-channel n-n potential in megaelectronovolts (MeV), as function of the distance in femtometers (fm) (inner axes) compared with the $O_2$-$O_2$ molecular potential in millielectronovolts (meV), as function of the distance in nanometers (nm) (outer axes).

Let me illustrate the main ideas of this approach by a didactic comparison between the neutron-neutron (n-n) potential (Argonne $v_{18}$) and the molecule-molecule $O_2$-$O_2$ potential in the textbook Lennard-Jones form. As shown in Fig. 3, a proper adjustment of the energy and distance scales makes both potentials fairly similar (the n-n potential is softer!). Both potentials have a mild long-range attraction and a strong short-range repulsion, neither of them binds the constituents, and both perfectly well describe low-energy scattering. In both cases the interacting objects are neutral: the $O_2$ molecule has neither QED charge nor dipole moment, and the neutron is QCD color white.

We understand perfectly well what happens when two $O_2$ molecules approach one another. At large distance, the electron clouds become polarized, and this induces dipole moments that generate the Van der Waals $1/r^6$ attraction. At small distance, very many things happen: higher polarization moments become important, direct Coulomb repulsion of electrons becomes important, and the Pauli blocking effects becomes important. All these things generate strong effects, and all can be modelled by a phenomenological $1/r^{12}$ potential, which has no real justification - it is just a repulsion. In fact, at low energies it is completely irrelevant what is the exact form of this repulsion. It can equally well be modelled by a proper contact interaction.

For the n-n system we still use the field-theoretical language when describing their long-range attraction, and we speak about one-pion and two-pion exchanges. I think that it could be extremely useful (at least on a pedagogical level) to retranslate these exchanges into the language of the QCD color polarization of neutrons, analogous to the QED $O_2$-$O_2$ case. In the molecular case nobody speaks about an exchange of a "particle" (a dipolon?) to describe the dipole-dipole Van der Waals attraction. We know of course that this force results from the QED photon exchanges, but who cares?

At small distance, the n-n color polarization becomes very complicated, and the Pauli blocking of valence quarks becomes also active. But when we probe neutrons within low-energy experiments, the details of all this are again completely irrelevant, and can be modelled by a properly adjusted contact force, as described above.

There is also another important lesson from the above simple comparison between the n-n and $O_2$-$O_2$ systems. Namely, when we keep two $O_2$ molecules at a fixed distance and the third one approaches, the first two become additionally polarized, and hence their interaction energy becomes modified. Hence, the interaction energy of three $O_2$ molecules is not a simple sum of three binary interactions – there must appear an explicit three-body term. If the analogy holds, one has to expect three-body NNN interactions between nucleons too.

This brings me to the talk of Witek Nazarewicz, who among other things told us about the recent progress in the exact calculations for low-energy states in light nuclei. There, the necessity of NNN interactions has been convincingly shown. Moreover, the NNN forces may also be responsible for a known inadequacy of the G-matrix method to derive the shell-model interactions. Witek also nicely discussed challenges of nuclear structure theory in describing exotic systems: like those having very large neutron or proton excess, very large angular momentum, or very large mass. In view of important projects to study these exotica in experiment (RIA, GSI, RIKEN, EURISOL, etc.), theoretical efforts in these domains must also be adequately expanded.

Two other talks, by Marek Płoszajczak and Krzysztof Rykaczewski, discussed other aspects of exotic nuclei. Marek introduced us to methods that combine advanced descriptions of bound nuclear states with equally advanced descriptions of scattering states. For weakly bound nuclei, such combined methods are essential. Unfortunately, they remained neglected for a (too) long time because of a necessity to expertly treat two fairly different physical situations. The so-called Gamow shell model has recently been devised to remedy this through a shell-model-like treatment of the particle continuum. Krzysztof, showed us that on the other side of the mass table, for proton unstable nuclei, the proton emission can be used as a fantastically efficient probe of nuclear states. By a careful analysis of proton radioactivity in deformed nuclei, we can explicitly see that the initial proton is really in a deformed state. This is one of the nicest examples of how the spontaneous symmetry breaking works in finite systems.

As a last item, I would like to mention the talk by Krzysztof Doroba who described recent advances in experimental verifications of the Standard Model. In particular, he described experiments that precisely measure mass, width, and other characteristics of the $Z$ and $W$ bosons – carriers of electroweak interactions. He convinced us that, once these basic physical constants are measured, one can within the Standard Model rigorously calculate very many things. I am always envious of successes of such exact theories, where everything can be, in principle, calculated with arbitrary precision. Nuclear physics, on the other hand, seems to be full of basically unsolvable problems. However, one can say that there are only two classes of problems in physics: unsolvable and trivial. We simply spend our lives on moving things from the first to the second category.

I would like to finish this talk by saying that it is my great pleasure and privilege to congratulate the organizers for an excellent result of their work. This has been really a great conference! I think we all enjoyed it very much and I hope we all meet here again in 2005.

The author acknowledges financial support from the Foundation for Polish Science (FNP) and from the Polish Committee for Scientific Research (KBN) under Contract No. 5 P03B 014 21.